\documentclass[11pt]{article}
\pdfoutput=1

\usepackage{euscript}
\usepackage{amssymb}
\usepackage{amsfonts}
\usepackage{amsbsy}
\usepackage{epsfig}
\usepackage{amsthm}
\usepackage{amscd}
\usepackage{amstext}
\usepackage{verbatim}
\usepackage{amsmath}
\usepackage{cancel}
\usepackage{capt-of}
\usepackage{empheq}
\usepackage{subfigure}

\usepackage[nosort]{cite}
\usepackage{bm}
\usepackage{authblk}
\usepackage{esint}
\usepackage[T1]{fontenc}
\usepackage{mathdots}
\usepackage[centertableaux]{ytableau}

\usepackage[utf8]{inputenc}
\usepackage{graphicx}
\usepackage{physics}
\usepackage{mathtools}
\usepackage{bbold}

\usepackage{setspace}
\usepackage{colortbl}
\usepackage{xcolor}

\usepackage[pdftex,linktoc=all]{hyperref}
\hypersetup{ 
colorlinks=true, 
linkcolor=black, 
citecolor=red, 
}

\def\ben{\begin{equation}}
\def\een{\end{equation}}
\def\half{{\textstyle{\frac{1}{2}}}}
\let\a=\alpha \let\b=\beta

\let\pa=\partial
\def\be{\begin{equation}}
\def\ee{\end{equation}}
\def\beq{\begin{equation}}
\def\eeq{\end{equation}}
\def\ba{\begin{array}}
\def\ea{\end{array}}

\def\dalemb#1#2{{\vbox{\hrule height .#2pt
       \hbox{\vrule width.#2pt height#1pt \kern#1pt
               \vrule width.#2pt}
       \hrule height.#2pt}}}

\newcommand{\bea}{\begin{eqnarray}}
\newcommand{\eea}{\end{eqnarray}}

\makeatletter
\newcommand*\bigcdot{\mathpalette\bigcdot@{.5}}
\newcommand*\bigcdot@[2]{\mathbin{\vcenter{\hbox{\scalebox{#2}{$\m@th#1\bullet$}}}}}
\makeatother

\renewcommand{\eqref}[1]{(\ref{#1})}

\title{Emergent Area Laws from Entangled Matrices}
\author{Alexander Frenkel$^\sharp$ and Sean A. Hartnoll$^{\flat}$}
\affil{\it $^\sharp$Department of Physics, Stanford University, \\
\it Stanford, CA 94305-4060, USA \\
\it $^\flat$Department of Applied Mathematics and Theoretical Physics, \\
\it University of Cambridge, Cambridge CB3 0WA, UK
}
\date{}

\begin{document}

\maketitle

\begin{abstract}

We consider a wavefunction of large $N$ matrices supported close to an emergent classical fuzzy sphere geometry. The $SU(N)$ Gauss law of the theory enforces correlations between the matrix degrees of freedom associated to a geometric subregion and their complement. We call this `Gauss law entanglement'. We show that the subregion degrees of freedom transform under a single dominant, low rank representation of $SU(N)$. The corresponding Gauss law entanglement entropy is given by the logarithm of the dimension of this dominant representation.
It is found that, after coarse-graining in momentum space, the $SU(N)$ Gauss law entanglement entropy is proportional to the geometric area bounding the subregion. The constant of proportionality goes like the inverse of an emergent Maxwell coupling constant, reminiscent of gravitational entropy.

\end{abstract}

\newpage

\tableofcontents

\section{Gauss law entanglement}

The quantum mechanics of large $N$ matrices is well-known to give rise to emergent space. The simplest example is the emergence of two dimensional string theory from the quantum mechanics of a single matrix  \cite{Klebanov:1991qa}, while the richest examples involve the emergence of critical string theory or M-theory from maximally supersymmetric Yang-Mills theories \cite{Maldacena:1997re} or quantum mechanics \cite{Banks:1996vh, Berenstein:2002jq}. There are many models in between these two limits, with varying numbers of matrices and degree of supersymmetry. A major open question is how, precisely, quantum states of matrices encode emergent geometry. In this paper we will make some general observations about the emergence of geometry from matrices, with specific computations in a bosonic model of three matrices recently discussed in \cite{Han:2019wue}.

We are able to make progress by restricting to technically tractable cases in which the geometry emerges from semiclassical matrices. This means that the matrix wavefunction is supported close to a classical noncommutative geometry, where the noncommutativity is small at large $N$. This is, of course, a major restriction. Such matrix states describe emergent $D$ brane worldvolumes \cite{Myers:1999ps} rather than gravitating spacetimes. In these cases we will show that the `areas' bounding subregions in the emergent space are directly encoded in what we call the Gauss law entanglement of the matrices. This is an entanglement due to the gauge-theoretic correlations in the matrix wavefunction, as we now explain.

Consider the matrix quantum mechanics of several $N$ by $N$ bosonic Hermitian matrices $X^{i}$ with conjugate momenta $\Pi^i$ and a Hamiltonian of the general form
\begin{equation}\label{eq:MQMlags}
    H = \Tr[\half \vec \Pi \cdot \vec \Pi + V(\vec X)] \,.
\end{equation}
The Hamiltonian is invariant under an $SU(N)$ symmetry generated by
\begin{equation}
    G = 2i\sum_i[X^{i},\Pi^i] \,.
\end{equation}
In a string theory context this symmetry is usually gauged, so that physical states $\ket{\Psi}$ obey:
\begin{equation}\label{eq:sing}
    G\ket{\Psi} = 0 \,.
\end{equation}
Even if the microscopic gauge constraint is relaxed \cite{Maldacena:2018vsr}, certain holographic states obey (\ref{eq:sing}).

The Gauss law (\ref{eq:sing}) expresses the global vanishing of $SU(N)$ charge. If one restricts attention to a subset of `local' degrees of freedom, these may carry a net charge. This charge is, of course, cancelled by the remaining degrees of freedom. Quantum mechanical fluctuations of the local charge will therefore result in, via the Gauss law, correlations between local degrees of freedom and their complement. This leads to what we call Gauss law entanglement. This is not a new concept, but we wish to give the phenomenon a name that emphasizes its physical origin. Gauss law entanglement in matrix theories was first discussed in \cite{Hampapura:2020hfg}. In more conventional local quantum systems it has been widely studied in the guise of gauge-theoretic `edge modes' \cite{Buividovich:2008gq, Donnelly:2011hn, Casini:2013rba,Ghosh:2015iwa,Donnelly:2016auv}. These modes are typically gauge transformations that do not vanish on the boundary of a geometric subregion and which therefore obstruct a geometric factorization of the gauge-invariant Hilbert space. Because the edge modes live on the boundary of subregions they produce `area' (or, better, `boundary') law entanglement.

It was shown in \cite{Frenkel:2021yql} that the $SU(N)$ Gauss law entanglement in the so-called Quantum Hall matrix model --- a matrix quantum mechanics with first order kinetic terms in the Lagrangian \cite{Susskind:2001fb, Polychronakos:2001mi} --- also exhibits a boundary law behavior, but now in an emergent two dimensional space. The Gauss law entanglement in the theory was shown to be in correspondence with that of large gauge transformations in an emergent Chern-Simons theory. Therefore, in that theory, matrix $SU(N)$ Gauss law correlations can be identified with conventional emergent geometric edge modes. However, the Quantum Hall matrix model leads to a topological gauge theory with no local dynamics in the emergent bulk. It is therefore at once rather simple while also containing subtleties specific to its topological nature. In this paper we will instead compute the Gauss law entanglement in a model of the form (\ref{eq:MQMlags}), with a second order kinetic term. The model is known to lead to an emergent $2+1$ dimensional Maxwell-scalar theory with genuine bulk dynamics on a fuzzy sphere \cite{Han:2019wue}.

Our main result is equation (\ref{eq:final}) for the $SU(N)$ Gauss law entanglement corresponding to a `cap' subregion of the emergent fuzzy sphere. This expression exhibits a geometric boundary law. Unlike in the Quantum Hall model discussed in the previous paragraph, however, this entanglement is not simply the edge mode entanglement of the emergent Maxwell field. The leading order matrix entanglement obtained in (\ref{eq:final}) is enhanced by a factor of the inverse dimensionless Maxwell coupling $\frac{\Lambda^{1/2}}{g_M}$. In this regard it has some similarities with the well-known $\frac{1}{G_N}$ contribution to gravitational entropy, as we discuss in \S\ref{sec:smear}. The matrix boundary law is furthermore enhanced by an unconventional logarithimic factor. The logarithm in (\ref{eq:final}) multiplies the boundary law and hence is distinct from the subleading universal logarithm widely discussed in Maxwell entanglement, e.g.~\cite{Donnelly:2015hxa, Radicevic:2015sza, Pretko:2015zva, Pretko:2018nsz}. We do not have a physical understanding of our logarithim; technically it arises as the asymptotic behavior of a simple counting problem in (\ref{eq:dimform}).

A further significant feature in our result (\ref{eq:final}) is the presence of a coarse-graining scale $\Lambda$. To obtain a geometric entropy it is essential to coarse-grain the reduced density matrix in order to smooth out the effects of UV/IR mixing in the noncommutative theory. This fact has implications for the general understanding of matrix entanglement. We initially define the subregion degrees of freedom via a block partition of matrices, as suggested by \cite{Das:2020jhy, Das:2020xoa}. However, the need for smearing to obtain a useful result shows that the block decomposition is `too microscopic' and is not, in and of itself, sufficient to reveal the emergent locality of the system. We are able to perform the smearing in our semi-classical state because we know a priori what the low energy field-theoretic degrees of freedom will be. For more strongly quantum mechanical states it may be difficult to make sense of the entropy following from the matrix block decomposition. Coarse-graining is discussed further in \S\ref{sec:smear}.

\begin{figure}[h]
    \centering
   \includegraphics[width=0.8\textwidth]{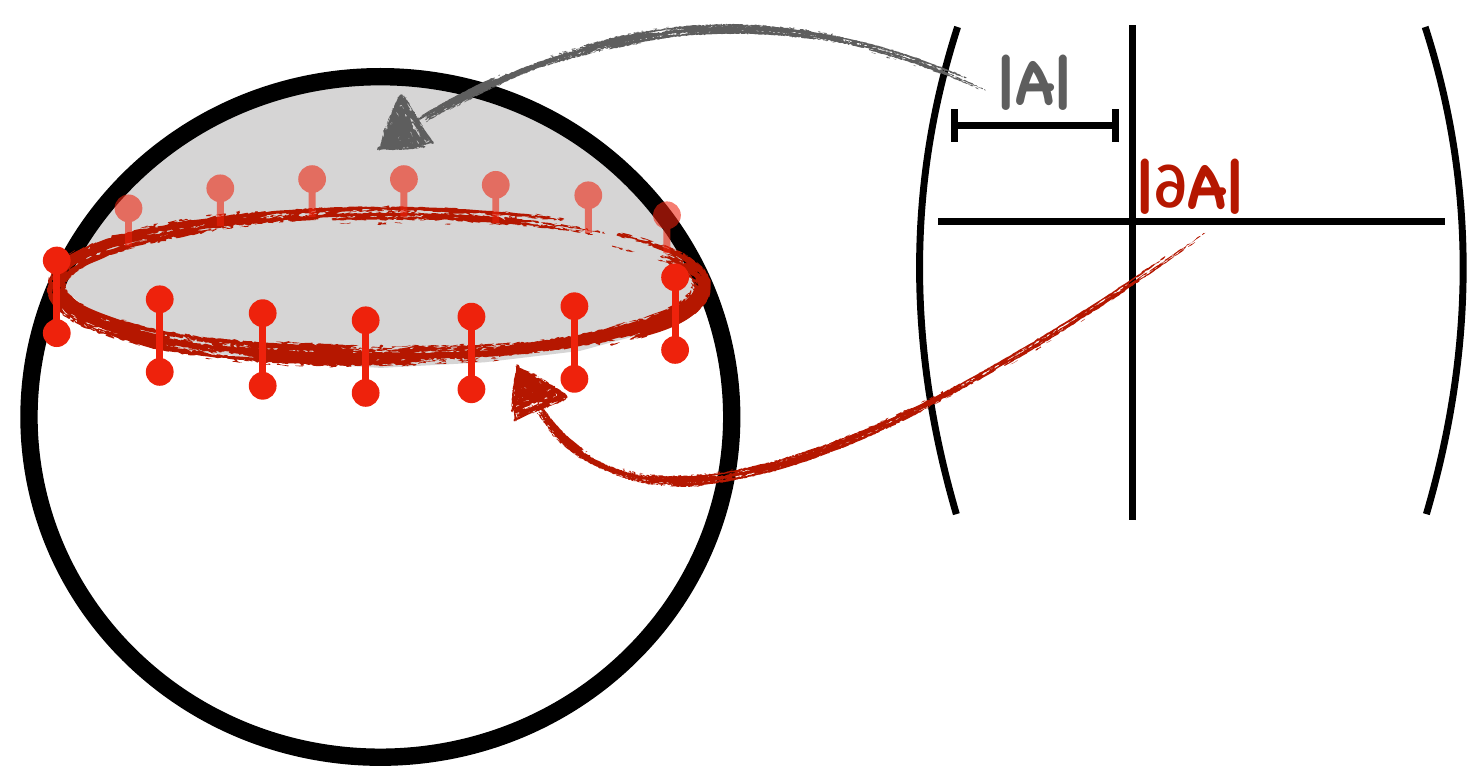}
    \caption{Geometric partition of the sphere and corresponding partition of the matrices. The `volume' (area) $|A|$ of the cap is given by the size of the upper left block of the matrix. The off-diagonal block of the matrices is low rank, with a unique non-vanishing singular value proportional to the `area' (perimeter) $|\partial A|$ of the cap. This singular value controls the $SU(N)$ charge of the cap and hence the Gauss law entanglement.}
    \label{fig:spheremat}
\end{figure}

While our computation will involve a few subtle points, the essential reason that a boundary law appears is simple and illustrated in Fig. \ref{fig:spheremat}. 
As we will recall in \S\ref{sec:project} shortly, a geometric subregion can be associated to a sub-block of the matrix degrees of freedom. In \S\ref{sec:gauge} we show, using the Gauss law, that the off-diagonal block determines the $SU(N)$ charge of the subregion. The off-diagonal block is found to be of low rank for states close to a classical non-commutative geometry, with a unique nonvanishing singular value given by the perimeter of the subregion --- this fact that also underpinned the emergence of an entanglement boundary law in the Quantum Hall matrix model \cite{Frenkel:2021yql}. This singular value will be seen in \S\ref{sec:low} and \S\ref{sec:cap} to determine the number of $SU(N)$ `edge modes' that are needed to ensure the Gauss law is obeyed by the reduced density matrix of the subregion. The $SU(N)$ edge modes will be shown to be maximally entangled and consequently lead to a boundary-law entanglement.

\section{Partitions of matrix systems}

\subsection{Projection onto subregions}
\label{sec:project}

We would like to partition the degrees of freedom in a matrix theory in a way that corresponds to emergent locality. In theories with a single matrix it is natural to partition the eigenvalues of the matrix. A `target space entanglement' \cite{Mazenc:2019ety} may be associated to this partition of eigenvalues, and has been shown to match the geometric entanglement of an emergent scalar field in a one dimensional space \cite{Das:1995vj, Das:1995jw, Hartnoll:2015fca}. Once there is more than one matrix in the theory, however, the various matrices cannot generically be simultaneously diagonalized.
On the other hand, recent works have noted that a given geometric partition in fact selects a single preferred matrix \cite{Das:2020jhy, Das:2020xoa}. The eigenvalues of this matrix may then be used to define a partition of all the matrix degrees of freedom, as we now briefly review.

In string theoretic matrix quantum mechanics there is commonly an association between the number of matrices and the number of emergent dimensions. That is, each matrix $X^i$ corresponds to an emergent coordinate $x^i$. In string theory the eigenvalues of the $X^i$ matrix give the location of $D$ branes in the $x^i$ direction. This correspondence suggests that subregions defined in terms of the coordinates can be pulled back to the matrix degrees of freedom. Specifically, suppose that a region $\Sigma$ in the emergent space is defined by $f_\Sigma(\vec x) < 0$. The function $f_\Sigma$ may be promoted to a function of matrices $f_\Sigma(\vec X)$. This step has an ordering ambiguity because the matrices do not commute. We will eventually coarse-grain the state over the noncommutativity scale, so this will not be a significant issue for us. The ordering ambiguity, however, already suggests the need to supplement the matrix block decomposition with a coarse-graining more generally. Given $f_\Sigma(\vec X)$, the crucial point is that $f_\Sigma(\vec X)$ is now a single matrix that may be diagonalized. For example, $f_\Sigma(\vec X) = X_1$ would correspond to a half-plane, while $f_\Sigma(\vec X) = \vec X \cdot \vec X - 1$ corresponds to the interior of a sphere. The eigenvalues of $f_\Sigma(\vec X)$ can be partitioned according to whether they are positive or negative, just as in the single-matrix theories mentioned above.

The partition of eigenvalues of $f_\Sigma$ is described in terms of a matrix and its complement:
\begin{equation}\label{eq:theta}
    \Theta_{\Sigma} \equiv \theta(f_\Sigma(\vec{X})), \qquad \Theta_{\bar{\Sigma}} \equiv \theta(-f_\Sigma(\vec{X})) = \mathbb{1} - \Theta_{\Sigma} \,.
\end{equation}
Here $\theta$ is the Heaviside step function. The matrix $\Theta_{\Sigma}$ is the matrix version of the characteristic function of the region $\Sigma$.
The eigenvalues of $\Theta_{\Sigma}$ are all 0 or 1 --- it is a projection matrix that obeys $\Theta_{\Sigma}^2 = \Theta_{\Sigma}$. The rank of $\Theta_\Sigma$ is defined to be $M$. With this projector at hand, all matrices may individually be split into four parts,
\begin{equation}\label{eq:matpart}
\left.
\begin{array}{rcl}
    X^{i}_{A B} & \equiv & \Theta_{A} X^{i} \, \Theta_{B} \\
    X^{i} & = & \sum_{A,B} X^{i}_{AB}
\end{array} \right\}
\qquad
A,B \in \{\Sigma, \bar{\Sigma}\} \,.
\end{equation}
The above construction is equivalent to going to a basis in which $f_\Sigma$ is diagonalized, with ordered eigenvalues. The projection $\Theta_\Sigma$ is onto the subspace spanned by the lowest $M$ eigenvectors of $f_\Sigma$. Equation (\ref{eq:matpart}) is block decomposition with respect to this subspace.

\subsection{Gauge charge in a subregion}
\label{sec:gauge}

We may now explain the interplay of the matrix partition defined in (\ref{eq:matpart}) with the Gauss law (\ref{eq:sing}). Related considerations have appeared previously in \cite{Hampapura:2020hfg, Frenkel:2021yql}. The most important conclusion of this subsection will be that the Gauss law leads to maximally mixed density matrices in each nontrivial charge sector of the `subregion'. This is equation (\ref{eq:max}) below.

The generator of the Gauss law (\ref{eq:sing}) may itself be partitioned using $\Theta_{\Sigma}$. This leads to the blocks $G_{\Sigma \Sigma}$, $G_{\Sigma \bar{\Sigma}}$, $G_{\bar \Sigma {\Sigma}}$ and $G_{\bar \Sigma \bar{\Sigma}}$, as in (\ref{eq:matpart}). It is clear that 
$G_{\Sigma \Sigma}$ are the $M^2$ generators of an $SU(M)$ subgroup of $SU(N)$, while $G_{\bar{\Sigma}\bar{\Sigma}}$
generate $SU(N-M)$. It is useful to write down these generators explicitly in terms of the partitioned matrix. In particular,
\begin{equation}
    G_{\Sigma \Sigma} = 2i\sum_{i}([X^i_{\Sigma \Sigma}, \Pi^i_{\Sigma \Sigma}] + X^i_{\Sigma \bar{\Sigma}}\Pi^i_{\bar{\Sigma} \Sigma} - \Pi^i_{\Sigma \bar{\Sigma}}X^i_{\bar{\Sigma} \Sigma}) \,.
\end{equation}
The singlet constraint (\ref{eq:sing}) then implies that
\begin{equation}\label{eq:aa}
    \sum_{i}[X^i_{\Sigma \Sigma}, \Pi^i_{\Sigma \Sigma}] \ket{\Psi} = \sum_i(\Pi^i_{\Sigma \bar{\Sigma}}X^i_{\bar{\Sigma} \Sigma} - X^i_{\Sigma \bar{\Sigma}}\Pi^i_{\bar{\Sigma} \Sigma}) \ket{\Psi}\,.
\end{equation}
We may re-interpret this expression as a Gauss law for a different $SU(M)$ that is generated purely by degrees of freedom within the region $\Sigma$. Namely, consider the $SU(M)$ generators
\begin{equation}
\hat G_{\Sigma\Sigma} \equiv 2i\sum_{i}[X^{i}_{\Sigma \Sigma},\Pi^{i}_{\Sigma \Sigma}] \,.
\end{equation}
We may further define charges made from off-diagonal degrees of freedom of the matrices
\begin{equation}\label{eq:QS}
Q_{\Sigma} \equiv 2i\sum_{i}\left( \Pi^{i}_{\Sigma \bar{\Sigma}}X^i_{\bar{\Sigma}\Sigma} - X^{i}_{\Sigma \bar{\Sigma}}\Pi^i_{\bar{\Sigma}\Sigma} \right) \,.
\end{equation}
The Gauss law (\ref{eq:aa}) becomes the statement that the off-diagonal degrees of freedom are charged under the $SU(M)$ generated by the $\hat G_{\Sigma\Sigma}$:
\begin{equation} \label{eq:g}
    \hat G_{\Sigma\Sigma} \ket{\Psi} = Q_{\Sigma} \ket{\Psi} \,.
\end{equation}

Tracing out the $\bar{\Sigma}\bar{\Sigma}$, $\bar{\Sigma}\Sigma$, and $\Sigma \bar{\Sigma}$ degrees of freedom leads to a reduced density matrix
\be
\rho_{\Sigma\Sigma} = \Tr_{\bar{\Sigma}\bar{\Sigma}, \bar{\Sigma}\Sigma, \Sigma \bar{\Sigma}} \ket{\Psi} \bra{\Psi} \,. 
\ee
Crucially this trace is over states in a larger `extended' Hilbert space in which the Gauss law is not imposed \cite{Buividovich:2008gq, Donnelly:2011hn, Casini:2013rba, Ghosh:2015iwa}. This is necessary because some of the $SU(N)$ transformations mix up the different blocks of the decomposition. Nonetheless, we may now use the fact that the state $\ket{\Psi}$ obeys (\ref{eq:g}) to obtain
\be\label{eq:terms}
[\hat G_{\Sigma\Sigma}, \rho_{\Sigma\Sigma}] = \Tr_{\bar{\Sigma}\bar{\Sigma}, \bar{\Sigma}\Sigma, \Sigma \bar{\Sigma}} \Big(Q_{\Sigma} \ket{\Psi} \bra{\Psi} - \ket{\Psi} \bra{\Psi} Q_{\Sigma} \Big) = 0\,.
\ee
The last step holds because $Q_{\Sigma}$ only acts on the degrees of freedom that are being traced over. Using the cyclic property of the trace, the two terms in (\ref{eq:terms}) cancel. Because all of the $M^2$ generators of $\hat G_{\Sigma\Sigma}$ commute with $\rho_{\Sigma\Sigma}$, the density matrix must be proportional to the identify on each nontrivial irreducible representation $R$ that appears in the Hilbert space.
That it is to say, the reduced density matrix for the gauge-theoretic `edge modes' is maximally mixed over each irreducible representation so that
\be\label{eq:max}
\rho_{\Sigma\Sigma} = \bigoplus_R \left(  \frac{p_R}{d_R} \mathbb{1}_{d_R} \bigotimes \rho_{R} \right) \,.
\ee
Here $d_R$ is the dimension of the representation, $p_R$ is the probability of the representation in the state and the $\rho_R$ are normalized density matrices in the unextended Hilbert space of the subregion.\footnote{More explicitly, let $\psi(X^i_{\Sigma\Sigma})$ be a wavefunction in the reduced Hilbert space of the region $\Sigma$. We may use an $SU(M)$ transformation $U$ to fix a gauge, for example by setting $X^i_{\Sigma\Sigma} = U X^{\text{gf}\,i}_{\Sigma\Sigma}  U^\dagger$, with $X^{\text{gf}\,3}_{\Sigma\Sigma} \equiv z$ diagonal and ordered. The wavefunction may then be written as 
\begin{equation}
    \ket{\psi} = \int dU dzdX_{\Sigma \Sigma}^{\text{gf}\,1} dX^{\text{gf}\,2}_{\Sigma \Sigma}\, \Psi(U,z,X_{\Sigma \Sigma}^{\text{gf}\,1},X_{\Sigma \Sigma}^{\text{gf}\,2})\,\ket{U} \otimes \ket{z,X^{\text{gf}\,1}_{\Sigma \Sigma},X^{\text{gf}\,2}_{\Sigma \Sigma}} \,.
\end{equation}
This defines a tensor factorization of the Hilbert space
$\mathcal{H}_{\Sigma\Sigma} = \mathcal{H}_U \otimes \mathcal{H}_\text{gf} \,,$
such that $SU(M)$ only acts on $\mathcal{H}_U$. From the Peter-Weyl theorem, then, $\mathcal{H}_{\Sigma\Sigma} = \left(\bigoplus_{R}\mathcal{H}_{R}\right) \otimes \mathcal{H}_\text{gf}$.\label{foot:one}
}

The entanglement entropy following from (\ref{eq:max}) is
\be\label{eq:Srho}
S(\rho_{\Sigma\Sigma}) = \sum_{R} p_R\log d_R - \sum_{R} p_R \log p_R + \sum_R p_R S(\rho_R) \,.
\ee
The first two terms here are what we are calling the Gauss law contribution to the entanglement. We will find that in our semiclassical model there is a single representation $R_o$ that dominates, so that $p_{R_o} \approx 1$ and all other probabilities are suppressed in the large $N$ limit. In this case the first term in (\ref{eq:Srho}) dominates and the Gauss law entanglement $S^\text{GL}$ is
\be\label{eq:Slog}
S^\text{GL}(\rho_{\Sigma\Sigma}) = \log d_{R_o} \,.
\ee
Computing $S^\text{GL}$ in this case therefore reduces to evaluating the dominant dimension $d_{R_o}$. The remaining final term in (\ref{eq:Srho}), from the unextended Hilbert space, is then $S(\rho_{R_o})$. We will compare the Gauss law and unextended contributions below.

The presence of Gauss law entanglement in matrix theories was first emphasized in \cite{Hampapura:2020hfg}, who focused on a $U(1)^{M-1}$ subgroup of $SU(M)$. We will see below that to obtain our leading order entanglement it is essential to consider the full $SU(M)$ `edge modes'. We had previously argued in \cite{Frenkel:2021yql} that the edge mode reduced density matrix should be maximally mixed over the dominant representation of $SU(M)$, but we believe the proof we have now given in (\ref{eq:terms}) is more direct.
It may be noted that there is some choice involved in embedding the physical state into an `ungauged' Hilbert space. Furthermore, as in the more familiar case of gauge theories \cite{Buividovich:2008gq, Donnelly:2011hn, Casini:2013rba,Ghosh:2015iwa,Donnelly:2016auv}, the Gauss law entanglement is due to $SU(M)$ edge modes that are introduced by the embedding (cf.~footnote \ref{foot:one}). The embedding is, however, a well-defined procedure that leads to a gauge-invariant Gauss law entanglement entropy that captures the tension between the Gauss law and locality.

\subsection{Low rank representations}
\label{sec:low}

In the remainder of this paper we will apply the formalism above to a state in a particular matrix quantum mechanics. This state will describe Gaussian fluctuations about a classical fuzzy sphere matrix configuration. We will see that acting on this state in the large $N$ limit, $\hat G_{\Sigma\Sigma}$ is dominated by a single representation $R_o$ and that in this representation $\hat G_{\Sigma\Sigma}$ is a low rank matrix. In fact $\hat G_{\Sigma\Sigma}$ will have rank two, which is the minimum possible rank for a nonzero traceless Hermitian matrix. We will now explain that in such cases there is a simple formula for the dimension $d_{R_o}$.

When $\hat G_{\Sigma\Sigma}$ is rank two it may be written in terms of a complex vector $\xi$ and its conjugate momentum $\pi$, so that
\be\label{eq:Qred}
(Q_\Sigma)_{ab} = i \left(\xi_b \pi_a - \xi^\dagger_a \pi^\dagger_b \right) \,.
\ee
Here $\xi$ will be a function of the matrix components of $X^i_{\Sigma \bar \Sigma}$ appearing in the charge (\ref{eq:QS}).
We have normal ordered the operators and $\pi_a = - i \pa/\pa\xi_a \equiv - i \pa_a$. The trace of the constraint (\ref{eq:aa}) implies that on physical states
\be\label{eq:Qcons}
\xi_a \pi_a |\Psi\rangle = \xi^\dagger_a \pi^\dagger_a |\Psi\rangle \,.
\ee
This constraint removes the $U(1)$ part of the generators (\ref{eq:Qred}). The generators (\ref{eq:Qred}) and constraint (\ref{eq:Qcons}) can be represented on polynomials in the $\xi_a$ of some fixed degree $\ell$ multiplied by degree $\ell$ polynomials in the complex conjugates $\bar \xi_a$. The dimension of the space of such polynomials is given by
\be\label{eq:drr}
d_R = \binom{M + \ell - 1}{M-1}^2 \,.
\ee
The square is due to the two independent polynomials. The product of polynomials does not immediately furnish an irreducible representation, because of the presence of singlet factors $(\xi \cdot \bar \xi)^\mu$ with $\mu \leq \ell$. However, the reducible polynomials are subleading at large $M$ and hence we may use (\ref{eq:drr}) as an approximation for the dimension of the dominant irreducible subspace. See \cite{Ahmadain:2022gfw} for a more group theoretic discussion of this point. We may further simplify (\ref{eq:drr}) when $M,\ell \gg 1$. In this limit, putting (\ref{eq:drr}) into the Gauss law entanglement entropy (\ref{eq:Slog}) gives
\be\label{eq:dimform}
S^\text{GL}(\rho_{\Sigma\Sigma}) = \log d_R \approx \begin{cases} 
\displaystyle 2 \ell \log \frac{\text{e} M}{\ell} & \ell \ll M \\[10pt]
\displaystyle 2 M \log \frac{\text{e} \, \ell}{M} & M \ll \ell   \end{cases} \,.
\ee

In order to evaluate (\ref{eq:dimform}) it remains to determine the appropriate value of $\ell$. In the remainder we do this in two steps. Firstly, we obtain the semiclassical quantum state of fluctuations about the fuzzy sphere. Secondly, we verify that upon restriction to a subregion the state transforms in a rank two representation, and we fix the dimension of the representation by computing the quadratic Casimir.

\section{The fuzzy sphere state}
\label{sec:fuzz}

\subsection{Classical fuzzy sphere}
\label{sec:fuzzclass}

We will focus on a model of three matrices $X^i$, $i=1,2,3$, with the potential in (\ref{eq:MQMlags}) given by
\be\label{eq:V}
V(\vec X) = \frac{1}{4} \left(\nu \epsilon^{ijk} X^k + i [X^i,X^j] \right)^2  \,.
\ee
Here $\nu$ is a mass deformation parameter. This model is the bosonic sector of a supersymmetric matrix quantum mechanics \cite{Asplund:2015yda}, which can be thought of as a `mini-BMN' theory \cite{Anous:2017mwr}. The potential energy in (\ref{eq:V}) is a total square. The classical ground states therefore obey the $SU(2)$ algebra
\be\label{eq:alg}
[X_\text{cl}^i, X_\text{cl}^j] = i \nu \epsilon^{ijk} X_\text{cl}^k \,.
\ee
We will be concerned with the maximal `fuzzy sphere' solution to (\ref{eq:alg}) given by
\be\label{eq:solX}
X_\text{cl}^i = \nu J^i \,,
\ee
with $J^i$ being the $N$ dimensional irreducible representation of $SU(2)$. Specifically, we take
\be\label{eq:solJ}
J^3 = \sum_{n=0}^{N-1} \left(\half (N-1)-n\right) |n)(n| \,, \quad
 J^+ = \sum_{n=1}^{N-1} \sqrt{n(N-n)} |n-1)(n|  \,, \quad J^- = (J^+)^\dagger \,,
\ee
where $J^\pm = J^1 \pm i J^2$.

The classical solution given by (\ref{eq:solX}) and (\ref{eq:solJ}) is just one of many degenerate minima of the potential, corresponding to different reducible representations of the algebra (\ref{eq:alg}). Domain walls between these different vacua were constructed in \cite{Bachas:2000dx}. Our objective here is to understand the emergence of geometry in a simple setting, and for this we restrict attention to a quantum state that is localized near the irreducible representation (\ref{eq:solJ}). We comment on other representations in \S\ref{sec:saddles}. The solution (\ref{eq:solX}) obeys
\be\label{eq:sphere}
(X_\text{cl}^1)^2 + (X_\text{cl}^2)^2 + (X_\text{cl}^3)^2 = \frac{\nu^2}{4} \left(N^2 - 1 \right) \,,
\ee
which is suggestive of a spherical geometry with radius $\sim \nu N/2$. In fact, the essential point about the fuzzy sphere solution is that fluctuations about the background can be organized in terms of matrix spherical harmonics. As we recall shortly, these are strongly analogous to the usual spherical harmonics, but are cut off at angular momentum $j_\text{max} = N - 1$. This angular momentum cutoff in the fluctuations is the sense in which (\ref{eq:solJ}) defines a {\it fuzzy} sphere. A smooth geometry is recovered in the limit $N \to \infty$.

\subsection{Fluctuations and semiclassical state}

We may now discuss quantum fluctuations about the classical solution. The semiclassical limit in which these fluctuations remain close to the classical geometry will be seen to be $\nu \to \infty$. It should be emphasized that this is logically distinct from the $N \to \infty$ limit in which the classical noncommutative geometry becomes smooth. We will be working in the limit in which both $\nu$ and $N$ are large.

Fluctuations about the fuzzy sphere can be written as
\be\label{eq:normal}
X^i = X_\text{cl}^i + \delta X^i = X_\text{cl}^i + 
\sum_a \delta x_a Y^i_a \,.
\ee
The normal modes $Y_a^i$ were obtained in \cite{Han:2019wue}, following \cite{Jatkar:2001uh, Dasgupta:2002hx}. We will review their construction in terms of matrix spherical harmonics shortly. The $Y_a^i$ should be taken to be Hermitian matrices. They obey the normalization condition $\sum_{i=1}^3 \tr \left[(Y^i_a)^\dagger Y^i_b \right] = \delta_{ab}\,,$
leading to the perturbed quadratic Hamiltonian
\be
H^{(2)} = \half \sum_a \left[\pi_a^2 +  \nu^2 \omega_a^2 (\delta x_a)^2 \right] \,.
\ee
Here $\pi_a$ is the momentum conjugate to $\delta x_a$. The normal mode frequencies $\omega_a$ will be given shortly. The (unnormalized) wavefunction close to the classical fuzzy sphere state is then
\be\label{eq:state}
\psi_\text{fs}(\delta x) = \prod_a \exp\left(- \half \nu |\omega_a| \delta x_a^2\right) \,.
\ee
The Gaussian wavefunction (\ref{eq:state}) is the semiclassical state from which we will obtain a Gauss law entanglement entropy.

It is apparent in (\ref{eq:state}) that at large $\nu$ the fluctuations about the classical solution are small. However, there are many normal modes at large $N$, so that their total zero point energy is large. This large zero point energy is cancelled in supersymmetric extensions of the theory, but in the purely bosonic theory it means that while (\ref{eq:state}) is a long-lived metastable state in the quantum theory --- and hence good enough as a tractable model of emergent geometry --- it is not the quantum ground state \cite{Han:2019wue}. We return to this point in \S\ref{sec:saddles}.

\subsection{The normal modes}

The entanglement depends on the structure of the normal modes in (\ref{eq:normal}). The modes can be expressed in terms of matrix spherical harmonics, a basis of matrices
$\hat{Y}_{jm}$ that obey
\begin{equation}\label{eq:JJJ}
    [J^3,\hat{Y}_{jm}] = m \hat{Y}_{jm}, \qquad \sum_{i}[J^i,[J^i,\hat{Y}_{jm}]] = j(j+1)\hat{Y}_{jm} \,.
\end{equation}
As with usual spherical harmonics, $-j \leq m \leq j$, but the angular momentum is now upper bounded by $j \leq N-1$. We review the construction of these matrices in Appendix \ref{sec:ylm}. In that Appendix we demonstrate that in the regime $N \gg |m|$ these matrices are given, in the basis used in (\ref{eq:solJ}) and for $m \geq 0$, by
\begin{equation}\label{eq:ylm}
    \hat{Y}_{jm} = 2 \sqrt{\pi} \, c_{jm} \sum_{k=1}^{N - m} Y_{jm}{\textstyle \left(\theta_k,0\right)}\, |k-1)(k+m-1| \,,
\end{equation}
where $Y_{jm}(\theta,\phi)$ are the usual spherical harmonics and the angles
\be\label{eq:cos}
\cos \theta_k \equiv 1 - \frac{2k + m}{N}\ \,.
\ee
Negative $m$ then follows from the relation $\hat{Y}_{j-m} = (-1)^m \hat{Y}_{jm}^{\dag}$. Equation (\ref{eq:ylm}) shows that, close to the commutative limit, the matrix spherical harmonics are essentially discretized spherical harmonics.

The overall prefactor in (\ref{eq:ylm}) is determined by
$\frac{1}{N} \Tr[\hat{Y}_{jm}^{\dag}\hat{Y}_{jm}] = \int_{S^2} d\Omega |Y_{jm}|^{2} = 1$. The coefficients $c_{jm}$ are given in Appendix \ref{sec:ylm}, the only property we shall need is that $c_{jm} \to 1$ to leading order with $N \gg m$.  We should note that various $m/N$ corrections are included in the above formulae: in the limit of the sum in (\ref{eq:ylm}), in the prefactor $c_{jm}$ and in the $m/N$ term in (\ref{eq:cos}). It will be physically illuminating to keep these corrections for the discussion of UV/IR mixing below, even while other $m/N$ corrections have been dropped --- see the discussion in Appendix \ref{sec:ylm}. Ultimately we shall truncate in \S\ref{sec:UVIR} to $j,m \leq \Lambda \ll N$, whereupon (\ref{eq:ylm}) is rigorously correct and the $m/N$ corrections are negligible.

The normal modes in (\ref{eq:normal}) may be expanded in terms of matrix spherical harmonics. A given mode may be written
\be\label{eq:modes}
Y^i = \sum_{jm} y^i_{jm} \hat{Y}_{jm} \,.
\ee
It will be convenient to introduce $y^\pm_{jm} = y^1_{jm} \pm i y^2_{jm}$.
In Appendix \ref{app:sols} we briefly review the equations for the coefficients $y^i_{jm}$, along with their explicit solution. The upshot is that the normal modes may be labelled by $j$ and for each $j$ there are three possible frequencies, each with a large degeneracy:
\begin{align}
    \omega & = 0  & \quad \text{multiplicity } N^2-1 \,, \nonumber \\
    \omega & =  - j & \quad \text{multiplicity } 2(j-1)+1 \,,  \label{eq:mid}\\
    \omega & = j + 1 & \quad \text{multiplicity } 2(j+1)+1 \,. \nonumber
\end{align}
Recall that $1 \leq j \leq N-1$. There are then $3 (N^2-1)$ normal modes in total. The zero modes are $SU(N)$ gauge transformations.
Because $\omega = 0$ for the global pure gauge modes, these do not appear in the wavefunction (\ref{eq:state}) and also will not contribute to any of our computations below. This confirms that the observables we compute using the gauge-fixed wavefunction are in fact gauge invariant. The remaining physical modes describe the coupled fluctuations of a Maxwell field and a scalar field on the fuzzy sphere \cite{Han:2019wue}. Each of the modes appearing in the above multiplets may be further labelled by $m$, as we describe in Appendix \ref{app:sols}.

\section{Cap subregion}
\label{sec:cap}

\subsection{$SU(N)$ charge}

With the quantum state at hand, we now wish to obtain the
charge $\ell$ of the subregion to use in the expression (\ref{eq:dimform}) for the Gauss law entanglement. We will do this by evaluating the expectation value of the quadratic Casimir $\left\langle \Tr (\hat G_{\Sigma \Sigma}^2) \right\rangle$ on the fuzzy sphere state. We will furthermore establish that the Casimir is strongly peaked on its expectation value and that the generators are low rank. That is to say, the state of the subregion is in a dominant low rank representation and fluctuations in the charge of the subregion are small.

On physical states the Gauss law (\ref{eq:g}) requires the generators of the reduced $SU(M)$ to be given by $\hat G_{\Sigma \Sigma} = Q_\Sigma$.
Inside the expectation value, we may expand the $SU(M)$ charge $Q_\Sigma$ given in (\ref{eq:QS}) around the classical configuration using the normal modes (\ref{eq:normal}) to obtain
\be\label{eq:Qexp}
Q_\Sigma \approx 2 i \nu \sum_i \sum_a \left(Y^i_{a \Sigma \bar \Sigma} J^i_{\bar \Sigma \Sigma} - J^i_{\Sigma \bar \Sigma} Y^i_{a \bar \Sigma \Sigma} \right) \delta \pi_a  \,.
\ee
Note that as an operator (\ref{eq:Qexp}) no longer obeys the $SU(M)$ algebra, as it has been linearized about its classical value. This linearization should be understood to be performed within the expectation value, and will be important later when we extract the value of $\ell$.

\begin{figure}[h]
    \centering
   \includegraphics[width=0.4\textwidth]{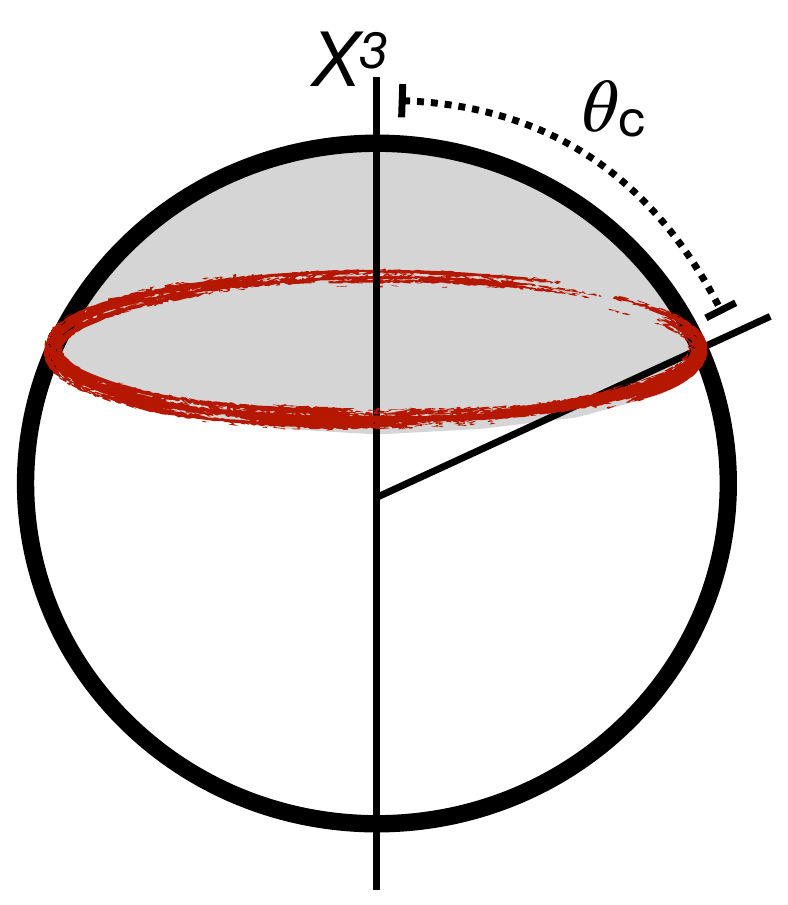}
    \caption{The cap subregion is shaded grey and is defined by the angle $\theta_c$ as shown.}
    \label{fig:angle}
\end{figure}

For simplicity we consider a `cap' subregion around the north pole of the fuzzy sphere. Such subregions are illustrated in Fig.~\ref{fig:angle}.
The cap is the region $\theta < \theta_\text{c}$ of the polar angle $\theta$. This region may equivalently be defined as a partition along the $X^3$ axis. Per our discussion in \S\ref{sec:project}, the $X^3$ direction is to be identified with the eigenvalues of $X^3_\text{cl}$. In (\ref{eq:solX}) and (\ref{eq:solJ}) we saw that $X^3_\text{cl}$ has eigenvalues equally spaced between $\nu N/2$ and $-\nu N/2$. We may project to the highest $M$ eigenvalues of $X^3_\text{cl}$ using
\be\label{eq:TM}
\Theta_\Sigma = \sum_{n=0}^{M-1} |n)(n| \,.
\ee
The projector has been written in the basis 
(\ref{eq:solJ}) in which $X^3_\text{cl}$ is diagonalized. This projection therefore corresponds to the region defined by the angle
\begin{equation}\label{eq:cosM}
    \cos \theta_c = 1 - \frac{2M}{N} \,.
\end{equation}
It follows from (\ref{eq:cosM}) that $M$ is indeed the `volume' (area) of the cap, as we stated in Fig.~\ref{fig:spheremat}. More precisely, in units where the sphere has unit radius, the area of the cap is $|A| = 4 \pi M/N$.

Using the projector (\ref{eq:TM}) we can see that the
matrix $Q_\Sigma = \hat G_{\Sigma \Sigma}$ in (\ref{eq:Qexp}) is low rank. Using the explicit form (\ref{eq:solJ}) of the $J^i$ gives
\be\label{eq:mat}
\left(\hat G_{\Sigma \Sigma}\right)_{PQ} = i \nu \sqrt{M(N-M)} \sum_a \left(Y^+_{aPM} \delta_{(M-1)Q} - \delta_{P(M-1)} Y^-_{a MQ} \right) \delta \pi_a \,.
\ee
Here $P$ and $Q$ run from $0$ to $M-1$. This matrix is manifestly of rank two and remains so at a quantum level because all of the $\delta \pi_a$ operators commute with each other. It may further be noted that the singular values of the matrix (\ref{eq:mat}) are proportional to
\be\label{eq:sin}
\sqrt{M(N-M)} = \frac{N \sin \theta_\text{c}}{2} \,.
\ee
This in turn is proportional to the length $|\partial A|$ of the boundary of the cap, giving a direct connection between the geometry of the subregion and gauge charge in the subregion. As in our previous work on matrix Chern-Simons theories \cite{Frenkel:2021yql}, this connection will be at the root of the emergent area law entanglement. This singular value is directly inherited from the off-diagonal block of the classical matrices in (\ref{eq:solJ}), as we illustrated in Fig.~\ref{fig:spheremat} above.

Using the explicit generators (\ref{eq:mat}) we may compute the quadratic Casimir. This will enable us to identify the representation in which the state transforms and thereby obtain the entanglement entropy via (\ref{eq:dimform}),
\begin{align}
\left\langle \Tr \left( \hat{G}_{\Sigma \Sigma}^2\right) \right\rangle_{\psi_\text{fs}} =
- \half\nu^3M(N-M)\sum_a |\omega_a| \left[ Y^{+\,2}_{a(M-1)M} + Y^{-\,2}_{aM(M-1)} - 2 \sum_{P=0}^{M-1} |Y^+_{aPM}|^2 \right]\label{eqn:trg2-line1}
\end{align}
Here we used the wavefunction (\ref{eq:state}), which implies $\left\langle \delta \pi_a \delta \pi_b \right\rangle_{\psi_\text{fs}} = \half \nu |\omega_a| \delta_{ab}$, and the classical background (\ref{eq:solJ}). We used $Y^+ = (Y^-)^\dagger$. The first two terms in \eqref{eqn:trg2-line1} may be dropped as they are subleading at large $N$. The remaining term may be evaluated using the formula for $Y^+$ given in Appendix \ref{app:sols}. As is explained in the Appendix, the different modes $a$ are labelled by $\{j,m\}$, with two families of modes (having $\omega=-j$ and $\omega=j+1$) for each $j$. The only fact we shall need from the Appendix is that for each mode
\be
Y^+ = \a_{jm\omega} \hat Y_{j(m+1)} + \b_{jm\omega} \hat Y_{j(m-1)}^\dagger \,, \qquad |\a_{jm\omega}|^2 + |\b_{jm\omega}|^2 = \frac{1}{N} \,.
\ee
Furthermore $\b_{j(-m)\omega} = (-1)^m\a_{jm\omega}$. We may then use the explicit large $N$ form (\ref{eq:ylm}) of the matrix spherical harmonics to obtain, being careful with the ranges of the various summations and to leading order at $j,|m| \gg 1$,
\be\label{eq:tg2}
\left\langle \Tr \left( \hat{G}_{\Sigma \Sigma}^2\right) \right\rangle_{\psi_\text{fs}} = 4 \pi \nu^3 \frac{M (N-M)}{N}\sum_{j=1}^N 2 j \sum_{m=1}^{\min(M,j)} c_{jm}^2 |Y_{jm}(\theta_{M-m})|^2 \,.
\ee
The above expression suggests the scaling $\Tr (\hat G_{\Sigma \Sigma}^2) \sim \nu^3 N^4$. Before developing (\ref{eq:tg2}) further, it remains to verify that quantum fluctuations in $\Tr (\hat G_{\Sigma \Sigma}^2)$ are small.
It is not uncommon for traces of matrices to become classical at large $N$, and in Appendix \ref{app:class} we verify that the variance of $\Tr (\hat G_{\Sigma \Sigma}^2)$ is indeed suppressed by a power of $1/N$. It follows that the reduced state is strongly peaked on a single low-rank representation of $SU(M)$.

\subsection{UV/IR mixing and smoothing}
\label{sec:UVIR}

We may make the geometric content of (\ref{eq:tg2}) clearer by re-expressing $M$ in terms of the angle $\theta_\text{c}$ of the cap in (\ref{eq:cosM}). In particular, using in addition (\ref{eq:sin}) and (\ref{eq:cos}), and changing the order of the summations, one obtains
\be\label{eq:tg22}
\left\langle \Tr \left( \hat{G}_{\Sigma \Sigma}^2\right) \right\rangle_{\psi_\text{fs}} = \pi \nu^3 N \sin^2\theta_\text{c} \sum_{m=1}^{M} \sum_{j=m}^N 2 j c_{jm}^2 |Y_{jm}(\cos^{-1}\left[\cos\theta_c + m/N \right])|^2 \,.
\ee
There are two aspects of the expression (\ref{eq:tg22}) that look unnatural from a continuum field theoretic perspective. The first is the $M$-dependent cutoff on the $m$ summation and the second is the shift by $m/N$ inside the inverse cosine.
As we now explain, these features of (\ref{eq:tg22}) are indeed not innocuous.

The noncommutativity of the fuzzy sphere coordinates in (\ref{eq:alg}) implies a (purely classical) uncertainty relation between any pair of coordinates. In particular, we may consider an excitation that is strongly localized to the sphere (\ref{eq:sphere}) of radius $R = \frac{\nu}{2} N$ while localized to a range $\Delta \theta$ about polar angle $\theta$ and a range $\Delta \phi$ about the azimuthal angle $\phi = 0$. Then 
\be\label{eq:thetaphi}
\Delta [\cos \theta] \Delta \phi = \frac{\Delta Z}{R} \frac{\Delta Y}{R \sin \theta} \geq \frac{\nu}{2} \frac{|\langle X \rangle|}{R^2 \sin \theta} = \frac{1}{N} \,. 
\ee
This equation explains both of the aforementioned unusual features in (\ref{eq:tg22}). For an excitation with azimuthal angular momentum $m$ we may put $\Delta \phi \sim 1/m$ in (\ref{eq:thetaphi}). It follows that $\Delta [\cos \theta] \gtrsim m/N$. A given azimuthal angular momentum therefore necessitates a smearing over the polar angle. This is precisely the effect seen in the argument of the spherical harmonic in (\ref{eq:tg22}). Secondly, from (\ref{eq:cosM}) we have $\Delta[\cos\theta] \sim \Delta M/N$. As we are considering modes that are restricted to a subregion with matrix size $M$, we must have $\Delta M \leq M$. The uncertainty relation therefore implies that $m \lesssim M$. This explains the upper cutoff in (\ref{eq:tg22}) of the sum over $m$ by $M$, which is stronger than the usual commutative bound of $m \leq j \leq N$.

The effects described in the previous paragraph are instances of UV/IR mixing, a phenomenon that is familiar in noncommutative field theories  \cite{Minwalla:1999px}. In particular, the bound $m \lesssim M$ is a short distance angular cutoff on modes around the boundary of a region that depends on the volume ($\propto M$) of the region enclosed. Given that (\ref{eq:tg22}) --- and hence, shortly, the entanglement entropy --- is dominated by short distance modes, it is clear that this volume-dependent cutoff will lead to violations of `boundary law' behavior. Similar effects have been noted in previous computations of the entanglement in noncommutative theories \cite{Karczmarek:2013xxa, Karczmarek:2013jca, Okuno:2015kuc, Chen:2017kfj}.
In fact, the expression is not even symmetric under $\theta_\text{c} \leftrightarrow \pi - \theta_\text{c}$.
This is possible because we have traced out the off-diagonal modes in $X^i_{\Sigma \bar \Sigma}$ and $X^i_{\bar \Sigma \Sigma}$ in addition to the complementary diagonal block $X^i_{\bar \Sigma \bar \Sigma}$. However, the asymmetry highlights the fact that (\ref{eq:tg22}) is non-geometric and does not encode a boundary law.

Recovering an approximately smooth geometry in the large $N$ limit requires smoothing over lengthscales that are afflicted by the UV/IR mixing that we have just described. The UV/IR mixing is significant because high angular momentum modes make a dominant contribution to (\ref{eq:tg22}). To reveal the approximate geometric locality contained in the matrix wavefunction we must coarse-grain the state.

The semiclassical matrix wavefunction (\ref{eq:state}) is a product of modes with different angular momenta $j$. It is therefore straightforward to trace out the high angular momentum modes in the full density matrix, by truncating
the angular momentum product over $j$ at some cutoff $\Lambda \ll M, N$. This leads to a smoothed density matrix $\rho^\Lambda \equiv |\psi^\Lambda_\text{fs}\rangle \langle \psi^\Lambda_\text{fs}|$. This tracing out of modes in a product state, that is done before tracing out the region $\bar \Sigma$, does not introduce any extra entanglement. In fact, the Gauss law entanglement will be strictly reduced as the high angular momentum modes are removed. Upon truncation (\ref{eq:tg22}) becomes
\begin{align}
\left\langle \Tr \left( \hat{G}_{\Sigma \Sigma}^2\right) \right\rangle_{\psi^\Lambda_\text{fs}} & = \pi \nu^3 N \sin^2\theta_\text{c} \sum_{j=1}^{\Lambda} \sum_{m=1}^{j} 2 j |Y_{jm}(\theta_c)|^2 \\
& \approx {\textstyle \frac{1}{6}} \nu^3 \Lambda^3 N \sin^2\theta_\text{c} \,. \label{eq:nice}
\end{align}
In the final expression, which is evaluated to leading order at large $\Lambda$, we used the fact that $\sum_{m=-j}^j |y_{jm}(\theta)|^2 = \frac{2j+1}{4\pi}$. We have also set $c_{jm}\to 1$ now that $m \ll N$.

In (\ref{eq:nice}) we see that in the smoothed state $\Tr (\hat G_{\Sigma \Sigma}^2) \propto |\partial A|^2$, the boundary `area' squared. We may now show that this implies a boundary-law entanglement entropy.

 \subsection{Gauss law entanglement entropy}

Having established that there is a dominant rank two representation of $SU(M)$, the quantity $\ell$ in the formula for the Gauss law entanglement entropy (\ref{eq:dimform}) may be obtained from the quadratic Casimir. For such representations of $SU(M)$,
$C_2 = 2 \left[\ell(M + \ell-1) - \frac{\ell^2}{M} \right]$ \cite{Gross:1993hu}, with the overall factor of $2$ because the edge mode states in \S\ref{sec:low} are the product of two polynomials. However, the linearization performed in (\ref{eq:Qexp}) means that we have not computed the full Casimir. Specifically, linearizing the generators directly rather than the Casimir itself means that we have effectively normal ordered the Casimir. From the low rank $SU(M)$ generators (\ref{eq:Qred}) one may show that $\langle (Q_\Sigma)_{ab} (Q_\Sigma)_{ba} \rangle = \langle :(Q_\Sigma)_{ab} (Q_\Sigma)_{ba}:\rangle + 2 \ell M$. That is, our linearized computation has missed the $2 \ell M$ contribution to the Casimir. The Casimir built from the linearized generators (\ref{eq:mat}) is simply a flat-space Laplacian type term, evaluating at large $\ell \gg 1$ to $C_2^\text{lin} \approx 2 \ell^2$. Therefore we have the identification
\be\label{eq:ell}
\ell \approx \left(\frac{1}{2} \left\langle \Tr \left( \hat{G}_{\Sigma \Sigma}^2\right) \right\rangle_{\psi^\Lambda_\text{fs}} \right)^{1/2} \approx \frac{(\nu \Lambda)^{3/2} N^{1/2}}{2\cdot 3^{1/2}} \sin \theta_\text{c} \,.
\ee

The entropy is obtained by using the value of $\ell$ from (\ref{eq:ell}) in the formula (\ref{eq:dimform}). In the limit that $\Lambda \ll \frac{1}{\nu} N^{1/3}$, which is weaker than the initial requirement that $\Lambda \ll N$ so long as $\nu$ is not extremely large, this leads to the coarse-grained Gauss law entanglement entropy associated to the cap:
\be\label{eq:final}
S^\text{GL}(\rho_{\Sigma\Sigma}^\Lambda) \approx \frac{1}{\sqrt{3 \pi}} \frac{\Lambda^{1/2}}{g_M} \frac{|\partial A|}{1/\Lambda} \times \log \frac{g_M N|A|}{\Lambda^{3/2} |\partial A|}  \,.
\ee
Here we have given the answer in terms of the length of the boundary of the spherical cap
\be
|\partial A| \equiv 2 \pi \sin \theta_\text{c} \,,
\ee
in units where the sphere has unit radius. Recall also that the volume of the cap $|A| = 4 \pi M/N$. We also introduced
the coupling constant of the non-commutative 2+1 dimensional Maxwell theory that arises on the emergent sphere \cite{Han:2019wue}
\be
g_M \equiv \sqrt{\frac{4 \pi}{N \nu^3}} \,.
\ee

Equation (\ref{eq:final}) is the promised boundary-law entanglement, up to a logarithmic correction. It is interesting to interpret (\ref{eq:final}) from the perspective of the low energy field theory on the emergent sphere. The term $\frac{|\partial A|}{1/\Lambda}$ is the length of the boundary region in units of a momentum cutoff $\Lambda$. This term is the natural magnitude for a conventional field-theoretic contribution to the entanglement. Consistently with this expectation, \cite{Han:2019wue} obtained $S \approx 0.03 \frac{|\partial A|}{1/\Lambda}$ for the entanglement in the unextended Hilbert space of this model. The contribution in \cite{Han:2019wue} should correspond to the final term in the full entanglement (\ref{eq:Srho}), albeit for a different microscopic partition of the degrees of freedom. The prefactor in (\ref{eq:final}), however, shows an enhancement by $\frac{\Lambda^{1/2}}{g_M}$. This is the inverse of the dimensionless coupling constant of the emergent theory and is large in the regime where the non-commutative Maxwell theory is weakly coupled.
Such enhancement of the gauge-theoretic entanglement due to microscopic degrees of freedom that are building the space itself may possibly be analogous to the large entropy found in string and gravitational theories \cite{Bekenstein:1973ur, Hawking:1975vcx, Susskind:1994sm, Ryu:2006bv}. The relationship between the microscopic and emergent field theoretic entanglement is discussed further in \S\ref{sec:smear}.

A further point to make is that there is a phase transition in the entanglement entropy if the cap becomes too small. At small angles $|A| \sim \theta_\text{c}^2$ and $|\pa A| \sim \theta_\text{c}$. For
\be
\theta_\text{c} \lesssim \frac{\Lambda^{3/2}}{N g_M} \,,
\ee
then the entropy (\ref{eq:dimform}) is to be evaluated in the regime $M \ll \ell$, leading to the `volume' law
\be
S^\text{GL}(\rho_{\Sigma\Sigma}^\Lambda) \approx \frac{N |A|}{2 \pi} \times \log \frac{\Lambda^{3/2} |\partial A|}{g_M N|A|} \,.
\ee
It should be emphasized that this transition is not due to UV/IR mixing effects, which have been smoothed out in our construction. Instead, it is due to a change in the asymptotic regime of the counting problem that determines the entropy. In this sense, the phase transition is similar in flavor to `deconfinement' type transitions in free gauge theories \cite{Aharony:2003sx}. It may be interesting to elaborate on this connection in the future.

\section{Discussion}
\label{sec:disc}

We have obtained the boundary-law entropy (\ref{eq:final}) as the Gauss law entanglement in a certain semiclassical state of a matrix quantum mechanics, associated to the matrix partition (\ref{eq:matpart}) together with a coarse-graining of momenta. The Gauss law entanglement captures the tension between the microscopic $SU(N)$ Gauss law constraint and the emergent locality in the theory. In the remainder we comment on some features of our construction that may warrant further thought, and also note possible connections to other discussions of entanglement and emergent geometry.

\subsection{Smearing and analogy with gravitational entropy}
\label{sec:smear}

To reveal the emergent locality it was necessary to supplement the matrix partition with a coarse-graining that smears over regions of size $1/\Lambda$. Without this smearing the dynamics is not local, due to the UV/IR mixing effects of spatial noncommutativity. The introduction of a smearing parameter $\Lambda$ means that the prefactor of the boundary-law entanglement (\ref{eq:final}) is not given in terms of the microscopic parameters of the model alone. It is plausible that this will be the case in any model with an emergent geometry: if it were possible to partition the microscopic degrees of freedom in a way that was in perfect correspondence with emergent low energy locality this would imply that the geometry is already present in the microscopic theory and not, in fact, emergent!

This point suggests that partitioning the matrix degrees of freedom as we have done, following the proposal in \cite{Das:2020jhy, Das:2020xoa}, is unlikely to be sufficient to capture geometric partitions. Additional information about the nature of the emergent geometry will be necessary. For semiclassical matrix states such as the one we have considered, the presence of a classical non-commutative geometry makes it possible to guess the correct smearing. However, in regimes where the matrices are strongly quantum the necessary smearing may not be obvious a priori. On the other hand, systems whose configuration space is just one matrix may be sufficiently simple that the microscopic partition is geometrically sensible \cite{Das:1995jw, Das:1995vj, Mazenc:2019ety, Frenkel:2021yql}. However, even in two-dimensional string theory there is a nonlocality in the map from the collective field to the string target space coordinates \cite{Moore:1991ag}.

We may note that the Bekenstein-Hawking-Ryu-Takayanagi contribution \cite{Bekenstein:1973ur, Hawking:1975vcx, Ryu:2006bv} to entropy in gravitational theories depends on Newton's constant, which is an effective coupling in the low energy theory and also not microscopic. The appearance of the dimensionless Maxwell coupling $\frac{\Lambda^{1/2}}{g_M}$ in our entropy (\ref{eq:final}) is somewhat reminiscent of that contribution and possibly should be thought of as an open-string analogue of the gravitational entropy. A further point in support of this analogy is that the entropy we have obtained can be thought of as counting strings that cross the entanglement cut, cf.~Fig.~\ref{fig:spheremat}, in the spirit of the proposal of Susskind and Uglum for gravitational entropy \cite{Susskind:1994sm}.

We may make two further remarks to emphasize that (\ref{eq:final}), while geometric, should be thought of as having a matrix rather than field-theoretic origin. The first is that the coupling $g_M$ is strictly a property of the non-commutative Maxwell theory, appearing in front of the intrinsically non-commutative term $[a_i,a_j]_\star$ in the $U(1)$ field strength \cite{Han:2019wue}. The second is that, as we recall in \S\ref{sec:symm} below, the emergent Maxwell gauge transformations are associated to a $U(1)^{M-1}$ subgroup of $SU(M)$. 
These are the diagonal components of the generators, in the basis where $X^3$ is diagonalized. However, as we now explain, this subgroup does not contribute to the result (\ref{eq:final}). From the generators (\ref{eq:mat}) ones sees immediately that the only nonzero diagonal component is $\left(\hat G_{\Sigma \Sigma}\right)_{(M-1)(M-1)}$. This means the only nontrivial edge modes transform under a  single $U(1) \subset U(1)^{M-1}$. However, $U(1)$ only has one dimensional representations and cannot contribute to the Gauss law entanglement.

The field-theoretic entanglement due to the emergent Maxwell and scalar field can instead to expected to arise at subleading order in the semiclassical expansion. The Casimir at the first subleading order is proportional, schematically, to $\langle (\delta x)^2 (\delta \pi)^2 \rangle$. Upon Wick contraction this quantity is independent of $\nu$, and hence of the Maxwell coupling $g_M$.

\subsection{Other saddles and the holographic direction}
\label{sec:saddles}

We have restricted attention to an especially simple  semiclassical state in the theory. As noted above, the state describes the polarization of many $D$0 branes into a single $D$2 brane, in the spirit of \cite{Myers:1999ps}, in a fixed background (for example, AdS$_4$ \cite{Asplund:2015yda}). In the regime studied, there is no backreaction of the branes onto the spacetime in the sense of AdS/CFT-type dualities \cite{Maldacena:1997re}. In particular, there does not seem to be an emergent `radial' or `holographic' dimension in the model. On the other hand, the emergent 2+1 dimensional Maxwell theory, which is coupled to a scalar, does have area preserving diffeomorphisms as a symmetry \cite{Han:2019wue}. This symmetry is inherited, see the following \S\ref{sec:symm}, from the original $SU(N)$ gauge invariance and the emergence of space via noncommuative geometry \cite{Lizzi:2001nd, Paniak:2002fi}. There may, therefore, be traces of gravity-like physics even in this simple state.

A more direct way to probe the holographic dimension is to consider other states of the theory, both semiclassical and beyond.
Within the semiclassical regime, reducible representations of the $SU(2)$ algebra (\ref{eq:alg}) give classical solutions describing multiple $D$2 branes. These can be coincident or separated. Coincident $D$2 branes will carry non-abelian gauge fields that at strong coupling will backreact into a holographic dimension. Distributions of $D2$ branes can also extend into the radial direction. An explicit realization of this connection arises in the maximally supersymmetric BMN matrix quantum mechanics \cite{Berenstein:2002jq}. In that theory all of the classical fuzzy sphere solutions survive as supersymmetric states in the quantum theory and are in correspondence with dual spacetime geometries \cite{Dasgupta:2002hx, Lin:2004nb, Lin:2005nh}.

We have already noted that the classical fuzzy sphere solution is not the ground state of the purely bosonic quantum theory.
Even in mini-BMN theory, the minimal supersymmetric completion of our model, with the same number of bosonic matrices, it was shown in \cite{Han:2019wue} that away from the semiclassical limit the fuzzy sphere state we have considered eventually collapses into a strongly quantum state that is not localized in the radial direction. That state is also a candidate for a more strongly holographic regime of the theory.

\subsection{Diffeomorphisms and permutations}
\label{sec:symm}

Several ingredients of our analysis have appeared previously in our computation of the entanglement in the Quantum Hall matrix model \cite{Frenkel:2021yql}. One important difference, however, concerns the role of permutations. In \cite{Frenkel:2021yql} we argued that states related by the symmetric group $S_M \subset SU(M)$ should not be considered as distinct 
edge modes. This changed the counting problem, as the edge modes were then given by symmetric polynomials rather than all polymonials of a given degree. In that context, the symmetrization appeared essential to recover the standard counting of chiral edge modes in Chern-Simons theory and the corresponding area law entanglement. In the present paper, however, we have found that an area law emerges without `re-gauging' the $S_M$ symmetry. We feel that a better understanding of this ambiguity is called for. In this subsection we explain that this is related to the role of diffomorphism invariance.

In the large $N$ limit, the $SU(N)$ symmetry group approaches the group of area preserving diffeomorphisms on the surface of the sphere \cite{Hoppe:1988gk, deWit:1988wri}, recently discussed in \cite{Donnelly:2022kfs}. See \cite{Han:2019wue} for an explicit discussion in our fuzzy sphere model. One way to see the connection to diffeomorphisms is to start by considering a particular diagonalized matrix to furnish a coordinate parameterization of the sphere. For example, in the bulk of this work we diagonalize $X^3 = \sum_n z_n v_n v_n^\dagger$. The eigenvectors $v_n$ define projectors $P_n \equiv v_n v_n^\dagger$ that have unit trace and are mutually orthogonal, and are therefore naturally associated (under the Moyal map) to non-overlapping unit area regions of the sphere. These regions are labelled by the eigenvalues $z_n$ which, in string theoretic realizations, are interpreted as the smeared locations of $D0$ branes. Under a unitary transformation each projector maps to a different projector $U P_n U^\dagger = v_n' v_n^{\prime\dagger} \equiv P_n'$. This transformation is therefore associated to a map from one set of non-overlapping unit area regions to another. This is the area preserving diffeomorphism.

Given the coordinate parameterization $X^3$ above, there are two subgroups of $SU(N)$ that do not change the associated set of unit area regions. The first are the $U(1)^{N-1}$ phases $U = \sum_n e^{i \theta_n} v_n v_n^\dagger$, which leave the regions invariant. 
As has been explained in \cite{Han:2019wue, Frenkel:2021yql}, these phase transformations are the origin of the emergent $U(1)$ Maxwell field in this model. Strings stretching between the $D$0 branes are charged under this transformation. The second set of special transformations are the permutations $S_N$ which exchange basis vectors and hence shuffle the regions among themselves. This amounts to a relabelling of the regions, which is not evidently a physical transformation. In \cite{Frenkel:2021yql} we found that quotienting the edge mode counting by these permutations lead to a sensible result for the entanglement entropy in the Quantum Hall matrix model. However, especially given the constraint (\ref{eq:Qcons}), this does not appear to be the case for the Gauss law entanglement in the fuzzy sphere model.

\pagebreak

\section*{Acknowledgements}

It is a pleasure to acknowledge helpful discussions with Jackson Fliss, Nabil Iqbal, Zohar Komargodski, John McGreevy and Ronak Soni.
We thank Juan Maldacena for correcting a formula in an earlier version of this paper.
The work of SAH is partially supported by Simons Investigator award \#620869 and by STFC consolidated grant ST/T000694/1. AF is
partially supported by the NSF GRFP under grant number DGE-165-6518 and by the US Department of Energy Office of Science under Award Number DE-SC0018134.

\appendix

\section{Form of the matrix spherical harmonics}
\label{sec:ylm}

In this appendix we derive expression (\ref{eq:ylm}) in the main text for the matrix spherical harmonics. We may start by recalling (see e.g.~\cite{Han:2019wue} for a recent discussion) that these matrices are constructed, in analogy with the usual spherical harmonics, via the generating relations
\begin{equation}
    \begin{split}
        &\hat{Y}_{j(-j)} = C(J^{-})^j \,,\\
        &\hat{Y}_{j(m+1)} = \frac{[J^{+},\hat{Y}_{jm}]}{\sqrt{(j-m)(j+1+m)}} \,.
    \end{split}\label{eq:build}
\end{equation}
The constant $C$ is to be fixed by the normalization $\tr \left(\hat{Y}_{j(-j)}^\dagger\hat{Y}_{j(-j)} \right) = N$.

Recall from (\ref{eq:solJ}) that the only nonzero entries of $J^{-}$ are $|k)(k-1|$, while $J^{+}$ has only $|k-1)(k|$ nonzero. One can then read off from (\ref{eq:build}) that the matrix $\hat{Y}_{jm}$ only has nonzero entries for elements $|k-1)(k+m-1|$. There is no approximation at this point. We now proceed to obtain an equation for these nonzero entries.
We set
\be\label{eq:Yform}
\hat{Y}_{jm} = \sum_{k=1}^{N-m} f(k) |k-1)(k+m-1| \,.
\ee
The objective is to solve for $f(k)$ in the large $N$ limit.

The matrix spherical harmonics are defined by the equations (\ref{eq:JJJ}). The first of these is automatically obeyed given the form (\ref{eq:Yform}). The second equation in (\ref{eq:JJJ}) is then seen to imply
\be\label{eq:eqjj}
[J^+,[J^-,\hat{Y}_{jm}]] = \Big( j(j+1)-m(m-1) \Big) \hat{Y}_{jm} \,.
\ee
In the large $N$ limit, the left hand side of this equation becomes a differential operator acting on $f(k)$. To see this, write
\begin{equation}
J^{+} = \sum_{k=1}^{N-1} g(k)|k-1)(k| \,.
\end{equation}
Recall also that $J^-$ is just the transpose of $J^+$. Then we have
\begin{align}
[J^+,[J^-,\hat{Y}_{jm}]] & = \Big(g(k+m-1) [g(k+m-1) f(k) - g(k-1) f(k-1)] \nonumber \\ & \qquad + g(k) [g(k)f(k) - g(k+m) f(k+1)] \Big)|k-1)(k+m-1| \label{eq:Delta}\\
& \approx  \Big(  -g(k)^2 f''(k) - 2 g(k) g'(k) f'(k) \nonumber \\
& \qquad + [m(m-1) g'(k)^2 - m g(k) g''(k)] f(k) \Big)|k-1)(k+m-1| \,. \label{eq:jpjm}
\end{align}
In the second line we have assumed that $k \gg \{m,1\}$ in order to Taylor expand the functions. If we use (\ref{eq:jpjm}) and (\ref{eq:Yform}) in equation (\ref{eq:eqjj}) we may equate the coefficients to obtain a differential equation for $f(k)$. Recall from (\ref{eq:solJ}) that $g(k) = \sqrt{k(N-k)}$. The differential equation acquires a familiar form if we perform the change of variables $k \equiv \frac{N}{2}(x+1)$, whereby
\be\label{eq:diffeq}
(1-x^2) f''(x) - 2 x f'(x) + \frac{m^2}{x^2-1} f(x)= - j(j+1) f(x) \,.
\ee
The regular solutions to this differential equation are the associated Legendre polynomials
\be\label{eq:ff}
f(x) = c_1 P_j^m(x) = c_2 Y_{jm}(\cos^{-1}(x),0) \,.
\ee
Here $c_1$ and $c_2$ are constants.

\begin{figure}[h!]
    \centering
   \includegraphics[width=0.6\textwidth]{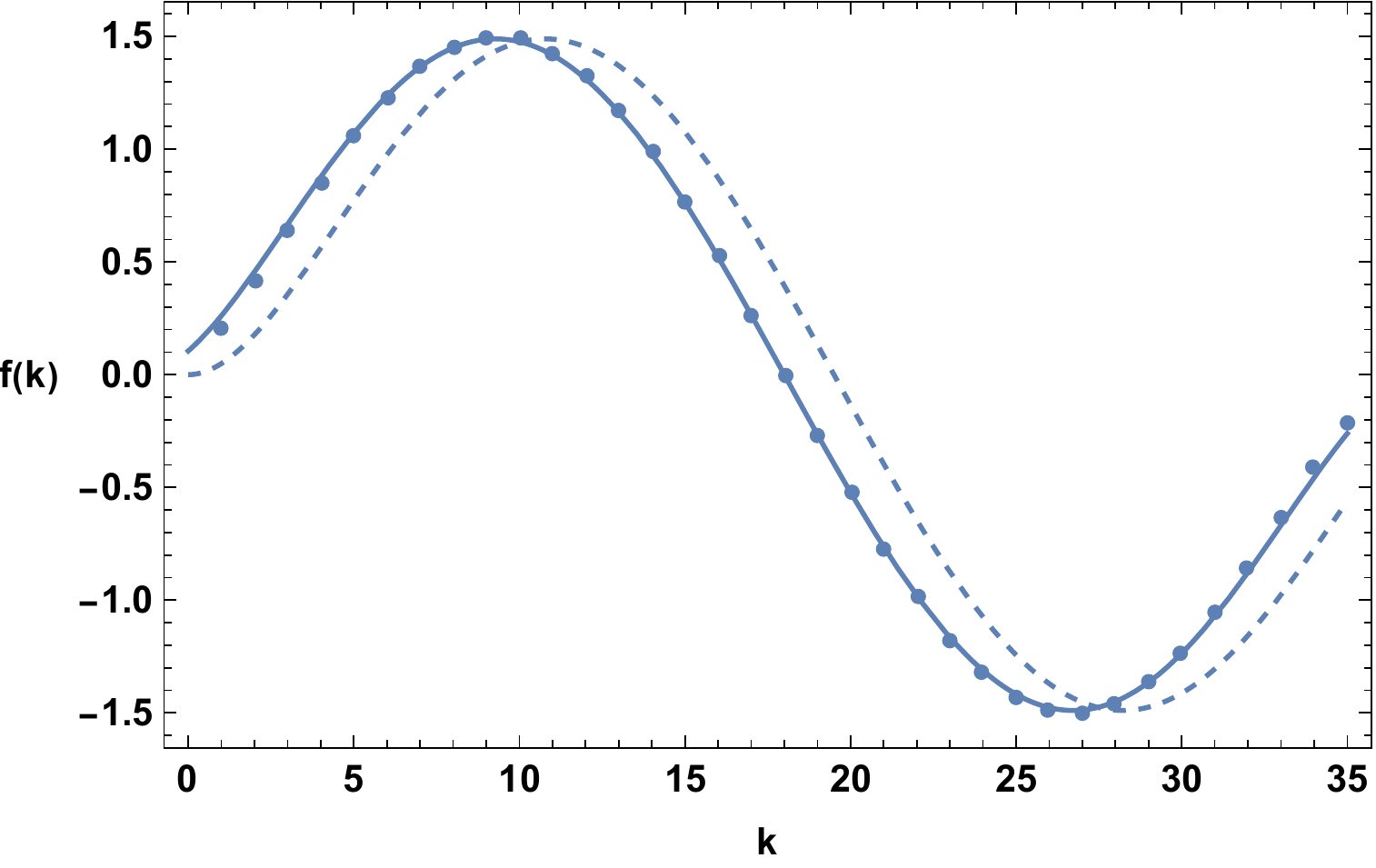}
    \caption{Comparison of the exact $f(k)$ in the matrix spherical harmonics (solid dots) with the expression (\ref{eq:shift}), given by the solid line, with $N = 39$, $j=5$ and $m=4$. The dashed line shows the expression (\ref{eq:shift}) without the shift by $(m-1)/N$. Even while small in this regime, the shift significantly improves the comparison.}
    \label{fig:mA}
\end{figure}

\begin{figure}[h!]
    \centering
   \includegraphics[width=0.6\textwidth]{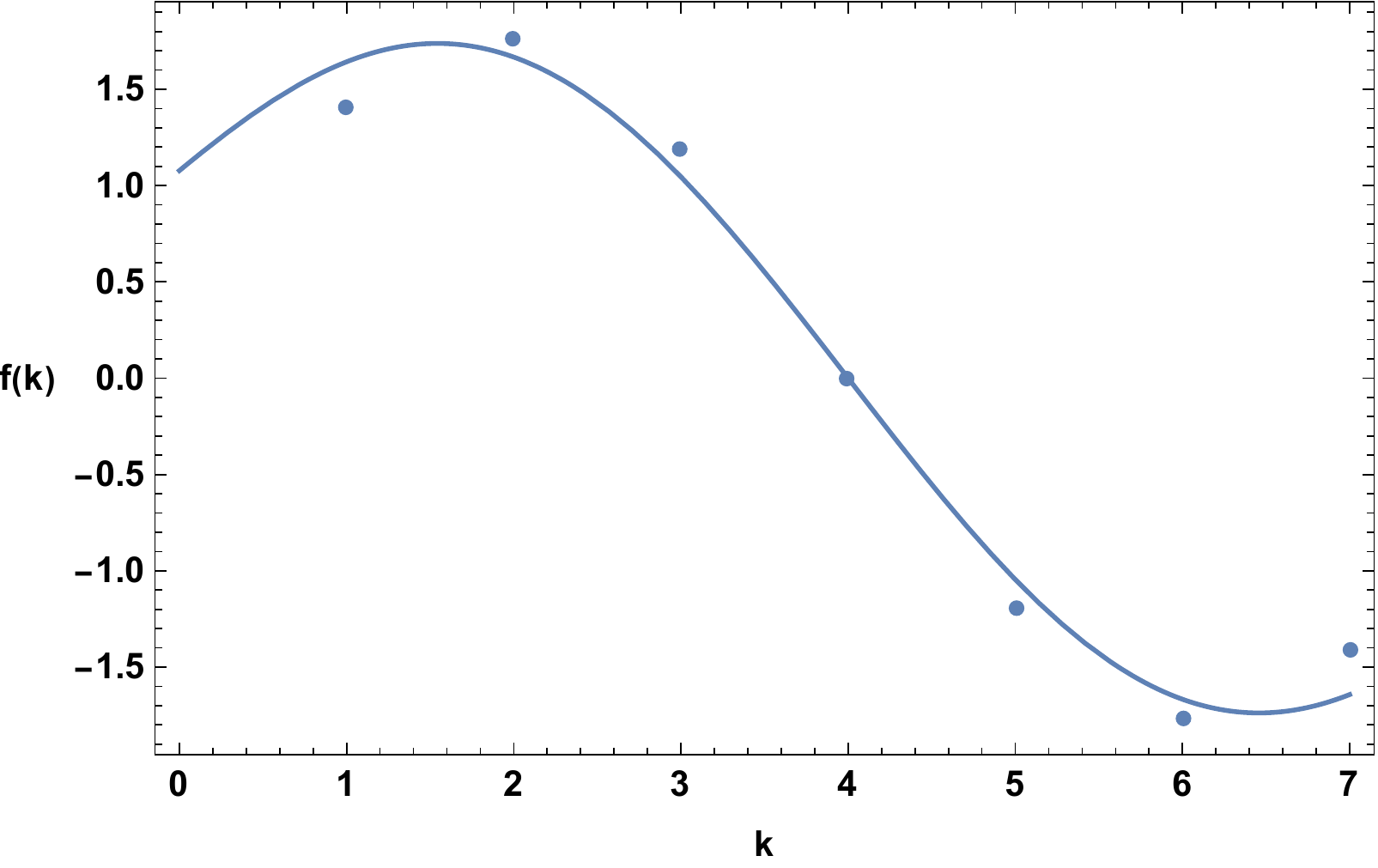}
    \caption{Same comparison as in Fig.~\ref{fig:mA} but now with $N=13$, $j=7$ and $m=6$. The agreement remains good despite noncommutative corrections being significant.}
    \label{fig:mB}
\end{figure}

Expression (\ref{eq:ff}) gives a direct connection between matrix spherical harmonics and regular spherical harmonics. This has required $k\sim N \gg m$. There are good reasons for this restriction. As we explain in \S\ref{sec:UVIR} in the main text, outside of this limit there are intrinsically matrix-like effects of spatial noncommutativity. However, an improvement to (\ref{eq:ff}) is possible that captures some of these effects. The full difference equation obeyed by $f(k)$, obtained from (\ref{eq:Delta}) and the explicit form for $g(k)$, is symmetric under the reflection $k \leftrightarrow N - m - k + 1$. Solutions for $f(k)$ can then be taken to be even or odd under this symmetry. The expression (\ref{eq:ff}) can be minimally promoted to one that manifests this symmetry by shifting the argument of the Legendre polynomial by $k \to k +(m-1)/2$ to give
\be\label{eq:shift}
f(k) = c_2 Y_{jm}\left(\cos^{-1}\left(1 - \frac{2k+m-1}{N}\right),0\right) \,.
\ee
This is the form we have quoted in the main text (\ref{eq:ylm}). The overall normalization $c_2$ is discussed in the following paragraph below. In (\ref{eq:cos}) in the main text we have dropped the final $-1/N$ term in (\ref{eq:shift}) that is subleading at large $N$. One may verify that while the shifted expression (\ref{eq:shift}) does remove some of the leading order $m/N$ corrections to the expanded equation of motion (\ref{eq:jpjm}), it does not remove all of them and therefore does not capture $m/N$ corrections in controlled way. Our final results in the main text, with the cutoff $\Lambda$ introduced, are only sensitive to the regime $m \ll N$ and do not depend on these corrections. The improved expression (\ref{eq:shift}) gives a sense of the effects of matrix noncommutativity --- see the discussion in \S\ref{sec:UVIR} --- and furthermore turns out to be in very good agreement with exact numerical answers even in the noncommutative regime, as illustrated in Figs.~\ref{fig:mA} and \ref{fig:mB}.

The normalization constant $c_2$ in (\ref{eq:shift}) is fixed by the normalization of the matrix spherical harmonics $\frac{1}{N} \Tr[\hat{Y}_{jm}^{\dag}\hat{Y}_{jm}] =1$ as well as the fact that $\int_{-1}^1 Y_{jm}(\cos^{-1}(x),0)^2 \, dx = (2\pi)^{-1}$. In the large $N$ limit the trace may be turned into an integral so that
\be
c_2^2 = 4 \pi c_{jm}^2 \,, \qquad \frac{1}{c_{jm}^2} = 2 \pi \int_{-1+ \frac{m}{N}}^{1 - \frac{m}{N}} Y_{jm}(\cos^{-1}(x),0)^2 \, dx
\ee

The sums over $j$ and $m$ that give the entanglement and other observables are dominated by large $j$ and $m$. Although we will not actually need to do so in this paper, for possible future reference we note that in this regime one may use the WKB form of the spherical harmonics
\begin{align}\label{eq:WKB}
Y_{jm}(\theta) = \frac{1}{\pi} \frac{\Theta(\sin^2\theta - \hat m^2)}{\left(\sin^2\theta - \hat m^2 \right)^{1/4}} & \times \\
\cos \Bigg(\frac{\pi}{4} - \Big(j + & \frac{1}{2} \Big) \left[\cos^{-1}\left(\frac{\cos\theta}{\sqrt{1 - \hat m^2}}\right) - \hat m \cos^{-1} \left(\frac{\hat m \cot\theta}{\sqrt{1 - \hat m^2}}\right) \right] \Bigg) \,, \nonumber
\end{align}
where $\hat m = m/(j + \frac{1}{2})$ and $\Theta$ is the heaviside step function. This form may be derived from the differential equation obeyed by the $Y_{jm}(\theta)$ and may be explicitly verified using numerics to match the full expressions very well at large $j$ and $m$.

\section{Details of the normal modes}
\label{app:sols}

The expansion (\ref{eq:modes}) for the normal modes (\ref{eq:normal}) leads to equations for the coefficients \cite{Han:2019wue}:
\begin{align}
        y_{jm}^3 + \frac{1}{2}\sqrt{(j+m+1)(j-m)}y_{j(m+1)}^+ - \frac{1}{2}\sqrt{(j-m-1)(j+m)}y_{j(m-1)}^- = \omega y_{jm}^3 \,, &\\
        (\omega \pm m)y_{j(m\pm1)}^{\pm} = \pm \sqrt{(j \pm m + 1)(j \mp m)}y_{jm}^3 \,. &\label{eq:eqs}
\end{align}
Here we see immediately that to find a mode with a given frequency $\omega$ we may pick a single nonzero $y_{jm}^3$, corresponding to a given value of $j$ and $m$, and then the only other coefficients that are forced to be nonzero from (\ref{eq:eqs}) are $y^+_{j(m+1)}$ and $y^-_{j(m-1)}$. 
For the perturbations $Y^i$ to be hermitian we must have
\be\label{eq:mmm}
    \bar{y}^{3}_{jm} = (-1)^m y^{3}_{j(-m)} \,, \qquad \bar{y}_{jm}^{+} = (-1)^m y_{j(-m)}^{-} \,,
\ee
These conditions further require that we have a nonzero $y_{j(-m)}^3$, $y^+_{j(-m+1)}$ and $y^-_{j(-m-1)}$.

In order to remain in the gauge slice that $X^3$ is diagonal, we must impose $Y^3 = 0$. To achieve this we may add a pure gauge mode, with $\omega = 0$, to the physical mode, which has $\omega \neq 0$. We may choose $y_{jm}^3(\omega) + y_{jm}^3(0) = 0$, so that the full mode physical mode has
\begin{align}
Y^3 & = \left(y_{jm}^3(\omega) + y_{jm}^3(0)\right) \hat Y_{jm} + \text{h.c.} = 0 \,, \label{eq:van} \\
Y^+ & = \left(y_{j(m+1)}^+(\omega) + y_{j(m+1)}^+(0)\right) \hat Y_{j(m+1)} + \left(\bar y_{j(m-1)}^-(\omega) + \bar y_{j(m-1)}^-(0)\right) \hat Y_{j(m-1)}^\dagger  \,. \label{eq:Ym}
\end{align}
And recall that $Y^+ = (Y^-)^\dagger$.
The normalization condition, given below (\ref{eq:normal}), implies that
\be\label{eq:nnn}
1 = \tr \left( (Y^+)^\dagger Y^+\right) = N \left(\left|y_{j(m+1)}^+(\omega) + y_{j(m+1)}^+(0)\right|^2 +  \left|y_{j(m-1)}^-(\omega) + y_{j(m-1)}^-(0)\right|^2  \right) \,.
\ee
Here we used the normalization of the matrix spherical harmonics, given below (\ref{eq:cos}), as well as the orthogonality of these harmonics at different values of $m$.

The equations (\ref{eq:eqs}), gauge condition (\ref{eq:van}) and normalization (\ref{eq:nnn}) may be solved explicitly to give the coefficients of the mode. For example,
\be
\left|y_{j(m-1)}^-(\omega) + y_{j(m-1)}^-(0)\right|^2 = \frac{1}{2N} \frac{(1+j-m)(j+m)(m+\omega)^2}{j^2(m^2+\omega^2) + j(m^2 + \omega^2)-m^2(m^2 + \omega(\omega-2))} \,.
\ee
This result determines the modulus of the first coefficient in (\ref{eq:Ym}). The phase of the coefficient can be chosen freely. Purely real and purely imaginary coefficients correspond to orthogonal Hermitian combinations of the $m$ and $-m$ modes.

\section{Classicality of the quadratic Casimir}
\label{app:class}

In this appendix we compute the variance of the quadratic Casimir $\Tr (\hat G_{\Sigma \Sigma}^2)$. The variance
may be obtained using the usual Wick contraction rules for the Gaussian wavefunction (\ref{eq:state}). In particular
\be
\left\langle \delta \pi_a \delta \pi_b \delta \pi_c \delta \pi_d \right\rangle_{\psi_\text{fs}} = \frac{\nu^2}{4} \left(|\omega_a| |\omega_c| \delta_{ab} \delta_{cd} + |\omega_a| |\omega_b|\delta_{ac} \delta_{bd} + |\omega_a| |\omega_b| \delta_{ad} \delta_{bc}\right) \,.
\ee
Therefore, using the expression (\ref{eq:mat}) for $\hat{G}_{\Sigma \Sigma}$ we obtain, to leading order at large $M,N$,
\begin{align}
\Big\langle \Tr \Big( & \hat{G}_{\Sigma \Sigma}^2\Big)^2 \Big\rangle_{\psi_\text{fs}} -
\left\langle \Tr \left( \hat{G}_{\Sigma \Sigma}^2\right) \right\rangle_{\psi_\text{fs}}^2 \\
& = \nu^6 M^2 (N-M)^2 \sum_{ab} |\omega_a| |\omega_b| \sum_{PQ} Y^-_{bMQ} Y^+_{aQM} \left( Y^-_{bMP} Y^+_{aPM} + Y^-_{aMP} Y^+_{bPM} \right) \\
& = 32 \pi^2 \nu^6 \frac{M^2 (N-M)^2}{N^2}\sum_{j,j'=1}^N 4 j j' \sum_{m=1}^{\min(M,j)} |Y_{jm}(\theta_{M-m})|^2 |Y_{j'm}(\theta_{M-m})|^2 \,. \label{eq:var}
\end{align}
The final expression follows from similar manipulations to those leading to (\ref{eq:tg2}) previously. We further used the fact that $Y_{jm}(\theta)$ is real. The variance (\ref{eq:var}) scales likes $\nu^6 N^7$. This is smaller than $\langle\Tr (\hat G_{\Sigma \Sigma}^2)\rangle^2 \sim \nu^6 N^8$ from (\ref{eq:tg2}). It follows that the large $N$ Casimir is classical.

\providecommand{\href}[2]{#2}\begingroup\raggedright\endgroup

\end{document}